\documentclass[twocolumn]{svjour3}
\smartqed
\usepackage{graphicx}
\usepackage{verbatim}
 \journalname{Granular Matter - DOI: 10.1007/s10035-009-0129-3}
\begin{document}

\title{Generation of Porous Particle Structures using the Void Expansion Method}

\author{Iwan Schenker \and Frank T. Filser \and Hans J. Herrmann \and \\
        Ludwig J. Gauckler
}

\institute{I. Schenker (corresponding author) \and F. T. Filser
\and L. J. Gauckler \at
              Nonmetallic Materials, Department of Materials, ETH Zurich, Zurich CH-8093, Switzerland\\
              \email{iwan.schenker@alumni.ethz.ch}           
           \and
           H. J. Herrmann \at
              Computational Physics for Engineering Materials, Institute for Building Materials, ETH Zurich, Zurich CH-8093, Switzerland
           }

\date{Received: 29. 6. 2008 / Accepted: 26. 1. 2009}

\sloppy

\maketitle

\begin{abstract}
The newly developed "void expansion method" allows for an
efficient generation of porous packings of spherical particles
over a wide range of volume fractions using the discrete element
method. Particles are randomly placed under addition of much
smaller "void-particles". Then, the void-particle radius is
increased repeatedly, thereby rearranging the structural particles
until formation of a dense particle packing.

The structural particles' mean coordination number was used to
characterize the evolving microstructures. At some void radius, a
transition from an initially low to a higher mean coordination
number is found, which was used to characterize the influence of
the various simulation parameters. For structural and
void-particle stiffnesses of the same order of magnitude, the
transition is found at constant total volume fraction slightly
below the random close packing limit. For decreasing void-particle
stiffness the transition is shifted towards a smaller
void-particle radius and becomes smoother.

\keywords{colloid \and coordination number \and discrete element
method \and microstructure generation \and porosity }
\end{abstract}

\section{Introduction} \label{intro}
Mechanical tests on coagulated colloids have shown that the local
arrangement of the colloidal particles has a strong influence on
the macroscopic mechanical properties. Colloids with a more
"heterogeneous" microstructure possess up to one order of
magnitude higher elastic modulii and yield strengths than their
"homogeneous" counterparts \cite{Ref1}.

Experimentally, colloidal microstructures with different degrees
of heterogeneity are obtained using an internal gelation method
(DCC = Direct Coagulation Casting \cite{Ref2,Ref3}). The method
allows for an in-situ, i.e. undisturbed, transition of the
inter-particle potential from repulsive to attractive. There are
two principal pathways leading to different microstructures:
changing the pH of the suspension ($\mathrm{\Delta}$pH-method) or
increasing the ionic strength in the suspension
($\mathrm{\Delta}$I-method). The first pathway shifts the pH to
the particles' isoelectric point and produces more "homogeneous"
microstructures through diffusion limited aggregation. In the
second pathway the ionic strength in the suspension is increased
at a constant pH which compresses the Debye length of the
repulsive potential leading to more "heterogeneous"
microstructures due to reaction rate limited aggregation of the
particles \cite{Ref4}.

Alternatively, heterogeneous microstructures can as well be
obtained by $\mathrm{\Delta}$pH-destabilization in conjunction
with small amounts of alkali-swellable polymer particles (ASP), 80
nm in diameter in the unswollen state \cite{Ref5}. The ASP
particles were admixed to the structural particles of 200 nm in
diameter under acidic conditions and swelled upon increasing pH
during the internal gelling reaction of the DCC process unfolding
to 800 nm in diameter, and thus pushing the structural particles
in their vicinity. Thereby, larger pores and thus more
heterogeneous microstructures are produced. Those samples with ASP
exhibit much higher mechanical properties than samples without
ASP. In particular, ASP samples present comparably high mechanical
properties as samples with heterogeneous microstructures produced
by the $\mathrm{\Delta}$I-method.

These experimental findings suggest that the colloid's
microstructure strongly determines its macroscopic mechanical
properties. The relation between structure and mechanical
properties, however, is not yet understood. One way to look at
this question is by computational means using simulation
techniques such as the discrete element method (DEM). This method
takes into account the particulate nature of a colloid and allows
for an investigation of the force distribution inside the particle
network during deformation as a function of the colloid's
microstructure. However, this method needs to be supplied with
initial particle configurations. In preceding works, Brownian
dynamics simulation (BD) was used to study the coagulation
dynamics and the evolving microstructures in colloidal suspensions
\cite{Ref6,Ref7}. These simulations were based on physical laws
and widely accepted theories such as the Stoke's drag force,
Brownian motion and the DLVO-theory \cite{Ref8}, describing the
inter-particle potential. The resulting microstructures agree well
with experiments \cite{Ref9} and can be used as initial particle
configurations for further DEM simulations to establish the link
between microstructure and macroscopic mechanical properties.
However, the BD method requires evaluating complex equations at
each time-step in order to determine the various forces acting on
the particles. Therefore, it is time-consuming, especially in the
case of a repulsive energy barrier and moreover, structures with
volume fractions exceeding 0.4 have not yet been simulated. For
processing reasons, ceramic engineers are interested in preferably
high solid's phase volume fractions and in particular in volume
fractions exceeding 0.4.

Inspired by the generation of heterogeneous micro-structures using
ASP we developed the "void expansion method" (VEM), which allows
for a fast and efficient computational generation of porous
microstructures over a broad range of volume fractions and
especially those exceeding 0.4.

In this publication, VEM is presented and the influence of various
simulation parameters such as the system size, the number of
particles within the system or the elastic properties of the
particles on the development of the mean coordination number is
analyzed for a wide range of volume fractions between 0.2 and
0.55.

\section{Materials and Methods}\label{sec:2}

\subsection{Discrete Element Method}\label{sec:21}

VEM is implemented using DEM \cite{Ref10} and in particular, the
particle flow code in three dimensions (PFC$^{\mathrm{3D}}$) from
Itasca Consulting Group, Inc., Minneapolis, Minnesota, USA
\cite{Ref11} is used. DEM is an iterative method in which discrete
spherical particles are used to build up more complex structures.
At each point in time the forces on each particle are calculated.
The time-step is chosen small enough to assume a constant force
during the time-step, which allows for the linearization of the
equations of motion enabling an efficient calculation of the
particles' next positions and velocities.

The forces on the particles included in our model arise from a
linear elastic contact law between the particles and damping. In
particular, no other forces such as long range forces between
particles or gravity are considered. PFC$^{\mathrm{3D}}$ uses a
soft-contact approach, wherein rigid particles are allowed to
overlap at contact points. The contact law relates the forces
acting on two contacting particles, in our case, linearly to the
relative displacement between these particles. The magnitude of
the normal contact force $F_n$ is given by Eq.~(\ref{eq:4})

\begin{equation}
 F_n = k_n U_n
 \label{eq:4}
\end{equation}

where $k_n$ denotes the normal stiffness and $U_n$ the overlap.
The shear stiffness $k_s$ relates an incremental displacement in
shear direction $\mathrm{\Delta} U_s$ to the shear contact force
$\mathrm{\Delta} F_s$ via Eq.~(\ref{eq:5}).

\begin{equation}
 \mathrm{\Delta} F_s = k_s \mathrm{\Delta} Us
 \label{eq:5}
\end{equation}

The linear elastic contact law is thus parameterized by its normal
and its shear particle stiffness.

Energy dissipation is introduced via a local damping term similar
to that described by Cundall \cite{Ref12}. The damping force,
characterized by its damping coefficient $d$, is added to the
equations of motion and is proportional to the force acting on the
particle. Thereby, only accelerating motion is damped and the
direction of the damping force is opposed to the particle's
velocity \cite{Ref11}.

Thus, the forces in our model are characterized by three
microscopic parameters: the particle's normal stiffness, its shear
stiffness and the damping coefficient. In this work the
inter-particle friction coefficient was set to zero in order to
allow the maximum particle rearrangement during the void
expansion.

\subsection{Void Expansion Method}\label{sec:22}

VEM relies on two distinct kinds of particles: "structural
particles" that constitute the final microstructure and
"void-particles" that are only used during the generation of the
structure. For clarity purposes, the first ones will be referred
to as \it structural particles \rm or simply particles and the
latter ones will explicitly be termed \it void-particles \rm
throughout this publication. The physically relevant macroscopic
parameters characterizing the final microstructures are $N_S$, the
number of structural particles, $r_S$, their radius and $\Phi_S$,
the volume fraction of the structural particles. Using these
parameters the edge length $l$ of the cubic simulation box with
periodic boundary conditions is calculated using Eq.~(\ref{eq:6}).

\begin{equation}
 l=r_s\left(\frac{4N_S\pi}{3\Phi_S}\right)^{1/3}
 \label{eq:6}
\end{equation}

The $N_S$ particles are randomly placed in the simulation box with
an initial particle radius of $r_S/(m+1)$, thus $(m+1)$ times
smaller than the final $r_S$, with $m$ being the number of
subsequent radius blow-up steps. We use $m = 10$ in our
simulations. At each blow-up step, the initial particle radius is
added to the current particle radius, followed by an equilibration
of the structure, until, after the $m^{\mathrm{th}}$ step, the
final particle radius $r_S$ is reached. This cyclic growing of the
particles is needed in order to achieve volume fractions higher
than approximately 0.35 without a considerable particle overlap
which represents high local stresses.

In addition to the structural particles, $N_V$ void-particles with
an initial radius $r_V \ll r_S$ are randomly placed in the
simulation box. We used $r_V = 0.005\,r_S$. After the structural
particles have reached their final size the radius of the
void-particles is increased cyclically. At each cycle their
initial radius is added to their current radius, thereby
simulating the swelling of the ASP. After each incremental
increase of the void-particle radius, relaxation steps are
performed in order to equilibrate the microstructure. This
iterative procedure is repeated until the structural and the
void-particles are densely packed and any further increase of the
void-particle size leads to a compaction of the particles, which
is reflected by an increase of the strain energy inside the
microstructure. Before each increase of the void particle radius
the positions of the structural particles are stored, which allows
for a subsequent analysis of the microstructure as function of
pore size, i.e. the void-particle's radius.

In this study, the mean coordination number $CN$ of the structural
particles alone is used to characterize the evolving
microstructures during the expansion of the void-particles. In
particular, the coordination number of a structural particle is
given by the number of neighboring structural particles with a
separation distance smaller than $d_{\epsilon} = (1 + \epsilon) 2
r_S $, with $\epsilon = 0.01$.

The density of bulk alumina was taken for the density of the
structural particles $\rho_S$. The void-particle density $\rho_V$
was set to a ten times smaller value $\rho_V = \rho_S/10$ in order
to reduce the inertia of the void-particles. Table~\ref{tab:1}
compiles the simulation parameters used in this work.

\begin{table}
\caption{Simulation parameters} \label{tab:1}
\begin{tabular}{lll}
\hline\noalign{\smallskip}
Parameter & Symbol & Value  \\
\noalign{\smallskip}\hline\noalign{\smallskip}
Number of particles & $N_S$ & 4000, 8000 \\
Particle radius & $r_S$ & 2.5 $\times$ 10$^{-7}$ m \\
Normal structural particle stiffness & $k_{n,S}$ & 10$^{2}$, 10$^{3}$ N/m \\
Shear structural particle stiffness & $k_{s,S}$ & 10$^{-3}$, 10$^{-2}$ N/m \\
Number of void-particles & $N_V$ & 400 - 16000 \\
Normal void-particle stiffness & $k_{n,V}$ & 10$^{-5}$ - 10$^{3}$ N/m \\
Shear void-particle stiffness & $k_{s,V}$ & 10$^{-9}$ - 10$^{-1}$ N/m \\
Damping coefficient & $d$ & 0.7 \\
Volume fraction & $\Phi_S$ & 0.2 - 0.55 \\
Structural particle density & $\rho_S$ & 3690 kg/m$^{3}$ \\
Void-particle density & $\rho_V$ & 369 kg/m$^{3}$ \\
\noalign{\smallskip}\hline
\end{tabular}
\end{table}

\section{Results and Discussion}\label{sec:3}

In this study, the mean coordination number $CN$ is used to
characterize the evolving microstructures during the expansion of
the void-particles. The evolution of $CN$ as function of the void-
to structural particle radius ratio $q = r_V/r_S$ is shown in Fig.
1 using 8000 structural particles, 2000 void-particles and a
volume fraction of 0.4. Also, the conventions of the nomenclatures
used throughout this paper are shown in this figure.

\begin{figure}
  \includegraphics{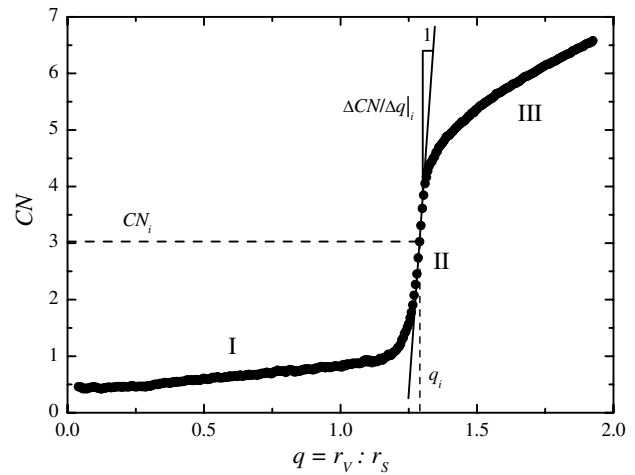}
\caption{Mean coordination number $CN$ as a function of the void-
to structural particle radius ratio $q$ using 8000 structural
particles, 2000 void-particles and a volume fraction of 0.4.}
\label{fig:figure1}
\end{figure}

The curve presents the three distinct regimes typical to all
curves analyzed throughout this study (Fig.~\ref{fig:figure1}):
The initial stage (I) is characterized by a small slope and a low
mean coordination number. The slope within the second stage (II)
increases drastically and the curve shows an inflection point. In
the third stage (III), the particles are densely packed and any
further increase of the coordination number is due to the
compaction of the particles reflected by a significant increase of
the structure's intrinsic strain energy. The transition stage can
be interpreted as a phase change between stage I, in which the
particles can move freely and stage III, in which the particles'
movements are arrested. The inflection point in the transition
region (II) is used to characterize the various curves that have
been simulated. It is defined by three parameters: the void- to
structural particle radius ratio $q_i = r_{V,i}/r_S$, with
$r_{V,i}$ being the void-particle radius at the inflection point,
the mean coordination number $CN_i$ and the maximum slope, denoted
by $\mathrm{\Delta}CN/\mathrm{\Delta}q\vert_i$. Further
characteristic parameters for the various curves are the void- to
structural particle number ratio $n = N_V/N_S$ and the targeted
volume fraction $\Phi_S$, which is the volume fraction of the
structural particles alone.

In the following, sensitivity analyses show the influence of
various VEM simulation parameters, especially the void- and
structural particle numbers (Sect.~\ref{sec:31}) and the targeted
volume fraction (Sect.~\ref{sec:32}), on $CN$. These simulations
use a void-particle normal and shear stiffness of 10$^{2}$ N/m and
10$^{-2}$ N/m, respectively and a structural particle normal and
shear stiffness of 10$^{3}$ N/m and 10$^{-2}$ N/m, respectively.
In Sect.~\ref{sec:33} the scaling behavior of $CN$ above the
inflection point as function of the total volume fraction is
analyzed. The influence of the void-particle stiffness on the
evolving microstructures is investigated thereafter in
Sect.~\ref{sec:34} for two distinct structural particle normal
stiffnesses: 10$^{3}$ N/m and 10$^{2}$ N/m. The structural
particles' normal to shear stiffness ratio was fixed at 10$^{5}$.

\subsection{Influence of the void- and structural particle numbers}\label{sec:31}

Microstructures with a volume fraction $\Phi_S = 0.4$ have been
generated using various numbers of structural and void-particles,
$N_S$ and $N_V$, respectively. $N_V$ essentially regulates the
void size. Indeed, the higher $N_V$ is chosen, the fewer blow-up
steps are necessary in order to densely pack the structural
particles. The influence of $N_V$ on the VEM was probed with $N_V$
ranging between 400 and 16000 for $N_S$ = 8000. The size
dependency of our system was tested with additional simulations
for $N_S$ = 4000 using 2000 and 4000 void-particles.

\begin{figure}
  \includegraphics{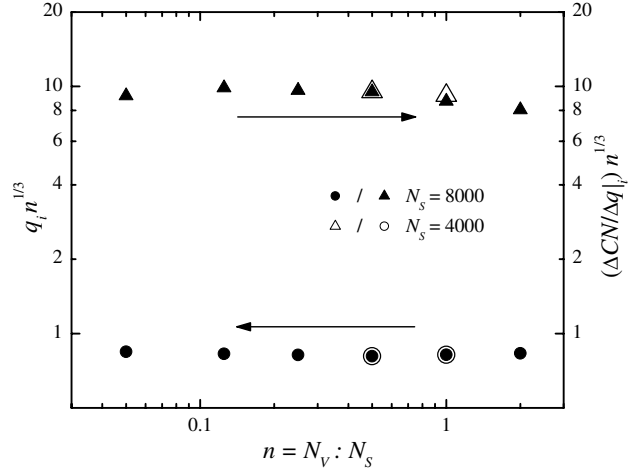}
\caption{Void- to structural particle radius ratio ($q_i$
normalized by $n^{-1/3}$) at the inflection point (circles, left
scale) and the corresponding differential increase in coordination
number ($\mathrm{\Delta}CN/\mathrm{\Delta}q\vert_i$ normalized by
$n^{-1/3}$, triangles, right scale) as a function of the void- to
structural particle number ratio ($n = N_V/N_S$). Open and filled
symbols denote simulations with $N_S$ = 4000 and 8000,
respectively.} \label{fig:figure2}
\end{figure}

The simulations show that the evolving microstructures depend on
the void- to structural particle number ratio $n$, but they are
independent on the individual absolute numbers of $N_V$ and $N_S$.
Increasing $n$ shifts the transition region (II) towards a smaller
void- to structural particle radius ratio $q$.
Fig.~\ref{fig:figure2} shows $q_i$ normalized by $n^{-1/3}$ as a
function of $n$ yielding a constant value given in
Eq.~(\ref{eq:7}).

\begin{equation}
 q_i n^{1/3} = 0.82 \pm 0.01
 \label{eq:7}
\end{equation}

The total volume fraction at the inflection point $\Phi_{T,i}$
(void and structural particles) is calculated via
Eq.~(\ref{eq:1}).

\begin{equation}
 \Phi_{T,i} = \Phi_{S}\left(1 + q_i^{3} n\right)
 \label{eq:1}
\end{equation}

Thus, a constant value of $q_i n^{1/3}$ entails a constant
$\Phi_{T,i}$. Indeed, the mean total volume fraction at the
inflection point is 62.3 $\pm$ 0.9 vol\% close to 64 vol\%, the
characteristic volume fraction of random close packings (RCP)
\cite{Ref13}.

Fig.~\ref{fig:figure2} further shows the differential increase in
coordination number at the inflection point
$\mathrm{\Delta}CN/\mathrm{\Delta}q\vert_i$ normalized by
$n^{-1/3}$ as a function of $n$ yielding a constant value given in
Eq.~(\ref{eq:8}).

\begin{equation}
 \mathrm{\Delta}CN/\mathrm{\Delta}q\vert_i n^{1/3} = 9.16 \pm 0.68
 \label{eq:8}
\end{equation}

The best power law fit of
$\mathrm{\Delta}CN/\mathrm{\Delta}q\vert_i$ as a function of $n$
yields an exponent of 0.3 slightly below 1/3 used for
normalization in Fig.~\ref{fig:figure2}. The fit is very good as
indicated by the correlation coefficient R$^2$ = 0.98.

In particular, the use of two distinct numbers of structural
particles ($N_S$ = 4000 and 8000) for two void- to structural
particle number ratios ($n$ = 0.5 and 1.0) shows that the
evolution of the mean coordination number is independent on the
system size. Virtually identical values for both $q_i n^{1/3}$ and
$\mathrm{\Delta}CN/\mathrm{\Delta}q\vert_i n^{1/3}$ as function of
$n$ were obtained as shown in Fig.~\ref{fig:figure2}.

\begin{figure}
  \includegraphics{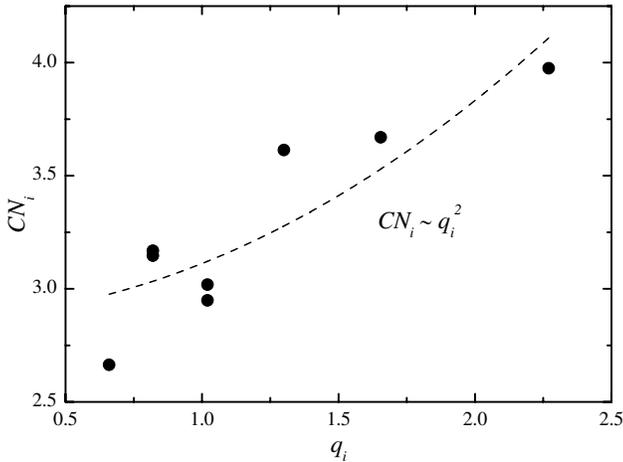}
\caption{Mean coordination number $CN_i$  as function of the void-
to structural particle radius ratio $q_i$, both at the inflection
point, for various void- to structural particle number ratios
$n$.} \label{fig:figure3}
\end{figure}

Combining Eq.~(\ref{eq:7}) and Eq.~(\ref{eq:8}) gives
Eq.~(\ref{eq:9})

\begin{equation}
 \mathrm{\Delta}CN\vert_i = 11.2q\mathrm{\Delta}q\vert_i
 \label{eq:9}
\end{equation}

which, after integration, results in Eq.~(\ref{eq:10})

\begin{equation}
 CN_i = 5.6q_i^2 + K
 \label{eq:10}
\end{equation}

with $K$ a constant. Thus, $CN_i$ is expected to scale as $q_i^2$,
which means that $CN_i$ essentially scales with the surface of the
void particles. $CN_i$ as a function of $q_i$ and its fit (dashed
line) are shown in Fig.~\ref{fig:figure3}. The trend is
reproduced, however, the quality of the fit is rather low (R$^2$ =
0.76). In particular, the two data points with $q_i$ close to
unity present a noticeable deviation of the general trend of
increasing $CN_i$ for increasing $q_i$. This case is remarkable as
it corresponds to an approximately monodispersed binary mixture of
the structural and void-particles. A more detailed investigation,
especially in the region of $q_i\approx1$, would require much more
simulation runs spanning a wider range of particle number ratios
$n$, which goes beyond this publication's scope.

\subsection{Influence of the volume fraction}\label{sec:32}

For processing reasons of ceramic bodies via colloidal routes
engineers are usually interested in volume fractions as high as
possible. A minimum of 40 vol\% was required to perform uniaxial
compression test on structures fabricated using the DCC process
\cite{Ref14}. Furthermore, the volume fraction is easily
accessible experimentally and is therefore widely used as a
comparative value for various experiments. Thus, the influence of
the volume fraction of the structural particles $\Phi_{S}$ was
investigated by means of $CN$ as a function of $q$ for $\Phi_{S}$
ranging from 0.2 to 0.55 and for $n$ = 0.5 ($N_S$ = 4000 and $N_V$
= 2000). The total volume fraction at the inflection point
$\Phi_{T,i}$ (void- and structural particles) is calculated using
Eq.~(\ref{eq:1}). Again, $\Phi_{T,i}$ yields a constant value of
61.5 $\pm$ 0.5 vol\%, slightly below the RCP limit. Thus,
$\Phi_{T,i}$ constitutes an upper boundary for $\Phi_{S}$. Indeed,
for $\Phi_{S}$ approaching $\Phi_{T,i}$ and $n \neq 0$, $q_i$
approaches zero and any expansion of the void-particles is
prohibited.

\begin{figure}
  \includegraphics{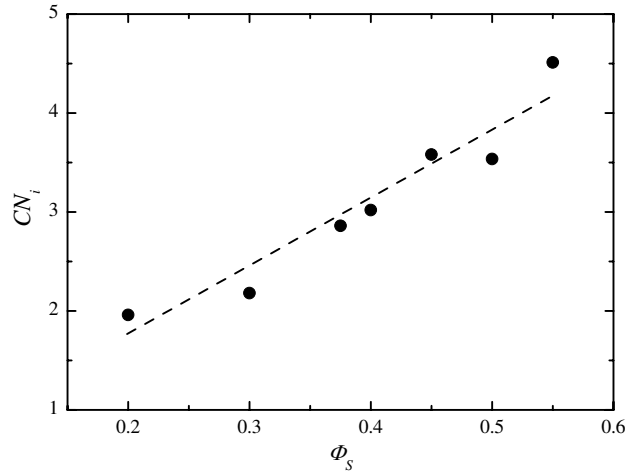}
\caption{Mean coordination number at the inflection point $CN_i$
as a function of the volume fraction of the structural particles
$\Phi_{S}$.} \label{fig:figure4}
\end{figure}

Fig.~\ref{fig:figure4} shows that $CN_i$ increases with increasing
volume fraction $\Phi_{S}$. As shown above, $\Phi_{T,i}$ is
constant for $\Phi_{S}$ ranging from 0.2 to 0.55, which indicates
that at the inflection point, the structural particles are equally
dense packed for all $\Phi_{S}$. Hence, $CN_i$ may be expected to
be a constant value, which seems to contradict
Fig.~\ref{fig:figure4}. Plausibility considerations based on
geometry give a possible explanation for this
$\Phi_{S}$-dependence of $CN_i$. The sum of $\Phi_{S}$ and
$\Phi_{V,i}$ equals $\Phi_{T,i}$, and is constant. Hence, rising
$\Phi_{S}$ results in a lower $\Phi_{V,i}$ and vice versa. Because
$N_V$ is constant, $\Phi_{V,i}$ only changes by the variation of
$r_{V,i}$. An increasing $\Phi_{V,i}$ entails an increase in
$r_{V,i}$, and vice versa. Structural particles in contact with
void-particles have a lower coordination number than those which
are only surrounded by other structural particles because only
contacts between structural particles are considered by definition
of $CN$. The number of structural particles in contact with
void-particles scales essentially with the total surface of the
void-particles. The smaller the total void surface, the less
contacts between structural and void-particles exist and the more
structural particles are only surrounded by other structural
particles. Hence, increasing $\Phi_{S}$ results in increasing
$CN_i$.

Similar considerations were used by Kruyt and Rothenburg
\cite{Ref15} who found a linear dependence between a particle's
coordination number and its radius in the case of two-dimensional
assemblies of polydisperse particles. Further studies are
necessary to confirm a linear dependence between $CN_i$ and
$\Phi_{S}$ as suggested by the line in Fig.~\ref{fig:figure4}.

\subsection{Scaling behavior of $CN$}\label{sec:33}

\begin{figure}
  \includegraphics{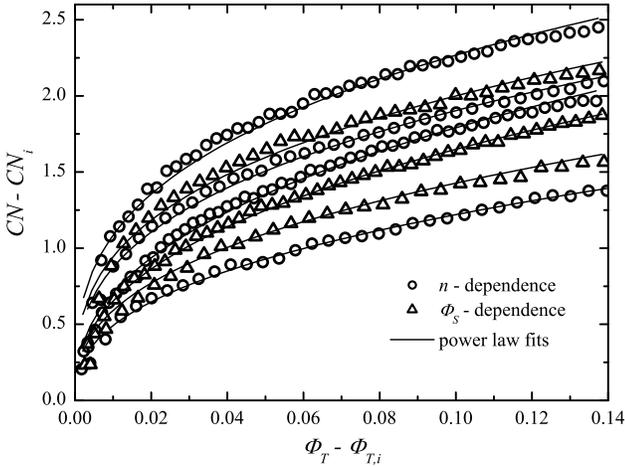}
\caption{Scaling behavior of the coordination number $CN$ above
the inflection point in dependence of the total volume fraction
$\Phi_T$ for various curves (open symbols) and corresponding power
law fits (lines).} \label{fig:figure5}
\end{figure}

In order to analyze the scaling behavior of $CN$ above $CN_i$
(region III) as a function of the total volume fraction $\Phi_T$
the various curves analyzed in Sect.~\ref{sec:31} and
Sect.~\ref{sec:32} were fitted using a power law given in
Eq.~(\ref{eq:13})

\begin{equation}
 (CN - CN_i) \propto (\Phi_T - \Phi_{T,i})^{\beta}
 \label{eq:13}
\end{equation}

with the exponent $\beta$ as fit parameter. A selection of these
curves is presented in Fig.~\ref{fig:figure5} for simulations as
function of the void- to structural particle number ratio $n$
(circles) and as function of the volume fraction of the structural
particles $\Phi_S$ (triangles). The corresponding power law fits
are shown as lines.

For the simulations with varying $n$, an exponent $\beta_{n} =
0.39 \pm 0.04$ was obtained. The simulations in dependence of
$\Phi_S$ yield $\beta_{\phi_S} = 0.35 \pm 0.03$. For all fits a
very high correlation coefficient R$^2 > 0.999$ was achieved
indicating excellent fits. In particular, our simulations suggest
that the exponents $\beta$ is independent of $n$ and $\Phi_S$. The
average over all simulations results in an exponent $\beta = 0.37
\pm 0.04$. This exponent is below the value 0.5 found in
literature \cite{Ref16,Ref17}, where however dense instead of
porous structures were considered.

\subsection{Influence of the void-particle stiffness}\label{sec:34}

\begin{figure}
  \includegraphics{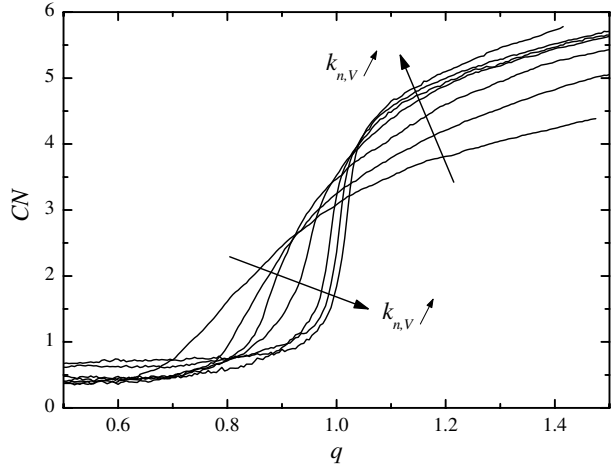}
\caption{Mean coordination number $CN$ as a function of the void-
to structural particle radius ratio $q$ for a void- to structural
particle number ratio $n = 0.5$ and for various void-particle
stiffnesses. The arrows indicate the direction of increasing
void-particle stiffness.} \label{fig:figure6}
\end{figure}

The void-particle stiffness essentially regulates the extent of
the overlap of a void-particle with structural particles or other
void-particles. Indeed, for a constant compressive force, a lower
void-particle stiffness allows for larger overlaps between a
void-particle and its neighbors. Thus, the void-particle stiffness
is an important parameter for the microstructure and its
evolution. For this investigation, the void-particle normal to
shear stiffness ratio $k_{n,V}/k_{s,V}$ is kept constant at
$10^4$, $n$ was fixed at 0.5 and $\Phi_{S}$ = 0.4.
Fig.~\ref{fig:figure6} presents the analysis of $CN$ as function
of $q$ for various values of $k_{n,V}$ ranging from 10$^{-5}$ N/m
to 10$^3$ N/m and for $k_{n,S} = 10^{3}$ N/m. Additionally, this
analysis was performed for $k_{n,S} = 10^{2}$ N/m and $k_{n,V}$
ranging from 10$^{-5}$ N/m to 10$^2$ N/m. In the following, the
results are presented as function of the dimensionless void- to
structural particle normal stiffness ratio $K_{n} =
k_{n,V}/k_{n,S}$.

Two principal behaviors are observed for increasing void-particle
stiffness and thus for increasing $K_{n}$: firstly, the inflection
point is continuously shifted towards larger void-particle sizes
and secondly, the transition between region I and III becomes
"sharper". These two observations will be elaborated in the
following.

\begin{figure}
  \includegraphics{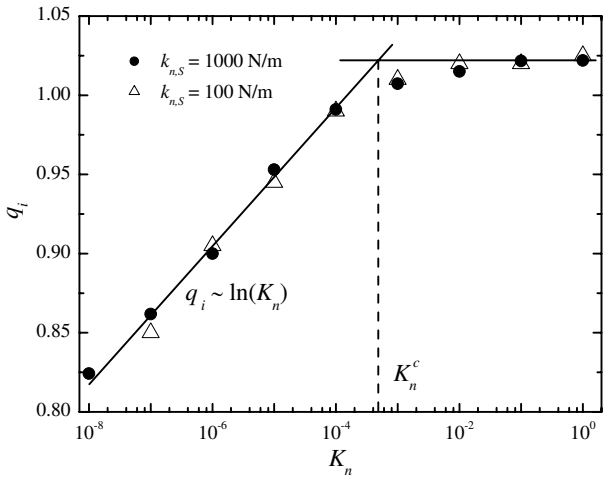}
\caption{Void- to structural particle radius ratio $q_i$ at the
inflection point as a function of the void- to structural particle
normal stiffness ratio $K_{n}$.} \label{fig:figure7}
\end{figure}

The shift of the inflection point towards larger void-particle
sizes for increasing $K_{n}$ is summarized in
Fig.~\ref{fig:figure7}, showing the continuous increase of $q_i$
for rising $K_{n}$. In particular, virtually identical curves are
obtained for the two distinct values of $k_{n,S}$. For small
normal stiffness ratios up to $K_{n} = 10^{-4}$ the values for
$q_i$ as function of $K_{n}$ follow a logarithmic law given in
Eq.~(\ref{eq:2}).

\begin{equation}
 q_i = 0.02 \ln K_n + 1.16
 \label{eq:2}
\end{equation}

The fit is very good as indicated by a correlation coefficient
R$^2$ = 0.993. For $K_{n} > 10^{-4}$, $q_i$ levels off at
approximately $q_i$ = 1.022, which corresponds to a total volume
fraction of 0.61, close to the RCP volume fraction value.

The characteristic void- to structural particle normal stiffness
ratio at which the transition from the logarithmic law to a
constant value occurs is given by the intersection point of the
respective fits. This value is termed $K_{n}^c$ and is found at
$5.4\times 10^{-4}$.

\begin{figure}
  \includegraphics{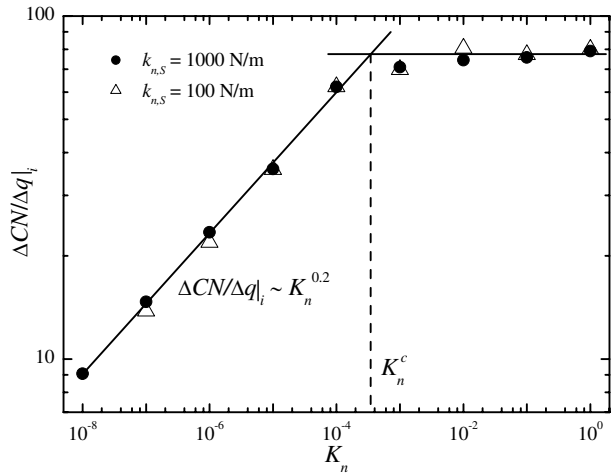}
\caption{Slope of the $CN(q)$-curves at the inflection point
$\mathrm{\Delta}CN/\mathrm{\Delta}q\vert_i$ as a function of the
void- to structural particle normal stiffness ratio $K_{n}$.}
\label{fig:figure8}
\end{figure}

A sharper, i.e. more "step"-like, transition between region I and
III is observed in Fig.~\ref{fig:figure6} for increasing
$k_{n,V}$. Mathematically, the "sharpness" of the transition
expresses in an increasing slope of the curves at their inflection
point, i.e. a higher value of
$\mathrm{\Delta}CN/\mathrm{\Delta}q\vert_i$, as summarized in
Fig.~\ref{fig:figure8} as function of $K_{n}$. As in the case of
$q_i(K_n)$ identical curves are obtained for the two values of
$k_{n,S}$. The values of
$\mathrm{\Delta}CN/\mathrm{\Delta}q\vert_i$ can be very well
fitted against the void- to structural particle normal stiffness
ratio $K_{n}$ up to $10^{-4}$ using a power law
(Eq.~(\ref{eq:3})).

\begin{equation}
 \mathrm{\Delta}CN/\mathrm{\Delta}q\vert_i = 410.3 K_{n}^{0.2}
 \label{eq:3}
\end{equation}

The correlation coefficient is R$^2$ = 0.997. For $K_{n}$-values
higher than $10^{-4}$ the
$\mathrm{\Delta}CN/\mathrm{\Delta}q\vert_i$-values level off.

The small exponent of 0.2 might also suggest a logarithmic
dependence between $\mathrm{\Delta}CN/\mathrm{\Delta}q\vert_i$ and
$K_n$, however, the logarithmic fit is of considerably lower
quality (R$^2$ = 0.91) compared to the power law fit.

The characteristic kink is found at $K_{n}^c = 3.5\times 10^{-4}$
N/m, which approximately corresponds to the value found for
$q_i(K_{n})$.

Eq.~(\ref{eq:2}) and Eq.~(\ref{eq:3}) describe the
$K_{n}$-dependence of $q_i$ and
$\mathrm{\Delta}CN/\mathrm{\Delta}q\vert_i$, respectively, for
$K_{n} \leq 10^{-4}$. These equations allow to express
$\mathrm{\Delta}CN/\mathrm{\Delta}q\vert_i$ as function of $q_i$
yielding a relation of the form
$\mathrm{\Delta}CN/\mathrm{\Delta}q\vert_i \sim \exp(q_i)$.
Indeed, an exponential fit gives a very high correlation
coefficient (R$^2$ = 0.992). The integration of
$\mathrm{\Delta}CN/\mathrm{\Delta}q\vert_i(q_i)$ allows to predict
the function $CN_i(q_i)$, which also results in an exponential
function. The fit of the simulated data using this function is
good as well, achieving a correlation coefficient of R$^2$ = 0.93.

\section{Summary and Conclusions}\label{sec:4}

In this paper we presented VEM which allows for an efficient and
fast computational generation of porous microstructures using DEM.
The development of VEM was inspired by the experimental generation
of heterogeneous colloidal microstructures using ASP. VEM is a
stochastic method in opposition to earlier used BD simulation, in
which the physical processes during the coagulation of the
colloidal suspension were simulated. This numerical description of
physical processes however requires much computing time. Thus,
from a computational viewpoint, VEM is less intensive than BD
simulation as VEM only includes a linear elastic contact law and
damping.

VEM permits to investigate the evolving microstructure as a
function of the void particle size. In order to characterize the
microstructure we used the mean coordination number. For all
simulation parameters used throughout this research, the mean
coordination number as function of the void- to structural
particle radius ratio exhibits the same characteristic
"step"-shape: a small slope with low mean coordination number in
an initial stage, a transition stage with a sharp increase of the
slope and, in a final stage, a small slope with high mean
coordination number. The transition can be seen as a phase change
between an initial stage, in which the particles can move freely
and a final stage, in which the particles' movements are arrested.
In this final stage, the particles are jammed and any further
swelling of the void-particles corresponds to an increase in the
strain energy in the structure. The inflection point in the
transition stage characterizes the obtained simulation curves and
enables a comparison between the various curves.

The sensitivity study of the mean coordination number by variation
of the void-particle number, the structural particle number, the
targeted volume fraction and the stiffness of the void-particles
leads to the following results:

\begin{enumerate}

\item The variation of the void- and structural particle numbers
and of the volume fraction of the structural particles reveals an
inflection point at a constant total volume fraction of
approximately 62\%, slightly below the RCP limit.

\item The total volume fraction of 62\% constitutes an upper
boundary for $\Phi_{S}$ and therefore for VEM.

\item The mean coordination number as function of the void- to
structural particle radius ratio depends on the void- to
structural particle number ratio alone and was found independent
of the system size.

\item Structures with volume fractions ranging from 0.2 to 0.55
were successfully simulated using VEM. In particular, volume
fractions above 0.4 were reached, which is crucial for a further
simulation of the uniaxial compression of colloidal structures
using DEM.

\item Above the inflection point, the mean coordination number as
function of the total volume fraction follows a power law with
exponent $\beta = 0.37 \pm 0.04$.

\item An increasing void- to structural particle stiffness ratio
$K_n$ reveals a twofold influence on the evolution of the mean
coordination number as a function of the void- to
structure-particle radius ratio: firstly, the inflection point is
shifted to higher void- to structural particle radius ratios and
secondly, the slope at the inflection point is increased.

\item For small $K_n$, the particle radius ratio $q_i$ and the
maximum slope $\mathrm{\Delta}CN/\mathrm{\Delta}q\vert_i$ at the
inflection point are nicely fitted versus $K_n$ using a power and
a logarithmic law, respectively. For $K_n$ approaching 1 the
curves level off. The transition from a power and logarithmic law,
respectively, to a constant value is found at a constant void- to
structural particle normal stiffness ratio $K_n^c$ of
approximately $4.5\times 10^{-4}$. In particular, the curves
$q_i(K_n)$ and $\mathrm{\Delta}CN/\mathrm{\Delta}q\vert_i(K_n)$ do
not depend on the structural particle stiffness.

\end{enumerate}

The computational time needed to generate a VEM structure is
determined by the number of void radius blow-up steps that are
necessary to densely pack the structural particles. The number of
void-particle blow-up steps essentially depends on $n$ and
$\Phi_{S}$ and is increasing for decreasing $n$ or $\Phi_{S}$.
Thus, smaller values of $n$ or $\Phi_{S}$ result in larger
computational times. The simulations have further shown that for
decreasing void-particle stiffness, less void blow-up steps are
needed in order to reach the inflection point. However, for
$k_{n,V} < 0.001 k_{n,S}$ lower packing density at the inflection
point are obtained for decreasing void-particle stiffness. In
PFC$^{\mathrm{3D}}$, the time-step essentially depends on the
particle's mass and its stiffness as $\sqrt{m_P/k}$, where $m_P$
is the smallest particle mass and $k$ the largest particle
stiffness in the system. Hence, for given particle densities the
time-step is determined by the structural particle normal
stiffness $k_{n,S}$ as long as $k_{n,V} \leq k_{n,S}$. For
$k_{n,V} > k_{n,S}$, the time-step decreases resulting in longer
simulation times.

To summarize, the VEM allows for an efficient computational
generation of porous microstructures over a wide range of volume
fractions and the various relations analyzed in this paper predict
the influence of the VEM simulation parameters on the
microstructures. Further research comprises the influence of
inter-particle friction, which in this study was set to zero in
order to facilitate at most any particle rearrangements during the
expansion of the void-particles.

We use VEM for an efficient generation of porous colloidal
microstructures over a wide range of volume fractions for
subsequent simulation of the mechanical properties using DEM.
Towards this goal, the VEM-microstructures have to withstand the
comparison with experimentally determined microstructures using
for example confocal laser microscopy \cite{Ref18} or with
structures obtained by other simulation techniques accurately
describing the physical processes during the coagulation such as
BD simulations \cite{Ref6,Ref7}. In this respect, an agreement in
the mean coordination number is a necessary but not a sufficient
condition. Additionally, further structural characterization
methods, such as the pair correlation function \cite{Ref6}, the
common neighbor distribution function \cite{Ref19} or the recently
introduced straight path distribution \cite{Ref20} must be
considered in order to quantitatively compare the microstructures
obtained by VEM to those obtained experimentally or using other
computational techniques.


\begin{thebibliography}{}

\bibitem{Ref1}
Wyss, H. M., Deliormanli, A. M., Tervoort, E., and Gauckler, L.
J., Influence of Microstructure on the Rheological Behavior of
Dense Particle Gels, AIChE J., 51 [\bf1\rm], 134-141 (2005)

\bibitem{Ref2}
Gauckler, L. J., Graule, Th., Baader, F., Ceramic forming using
enzyme catalyzed reactions, Mater. Chem. Phys., 61, 78-102 (1999)

\bibitem{Ref3}
Tervoort, E., Tervoort, T. A. and Gauckler, L. J., Chemical
Aspects of Direct Coagulation Casting of Alumina Suspensions, J.
Am. Ceram. Soc., 87 [\bf8\rm], 1530-1535 (2004)

\bibitem{Ref4}
Wyss, H. M., Tervoort, E., Meier, L. P., M\"uller, M. and
Gauckler, L. J., Relation between microstructure and mechanical
behavior of concentrated silica gels, J. Colloid Interface Sci.,
273, 455-462 (2004)

\bibitem{Ref5}
Hesselbarth, D., Tervoort, E., Urban, C. and Gauckler, L. J.,
Mechanical Properties of Coagulated Wet Particle Networks with
Alkali-Swellable Thickeners, J. Am. Ceram. Soc., 84 [\bf8\rm],
1689–1695 (2001)

\bibitem{Ref6}
H\"utter, M., Local Structure Evolution in Particle Network
Formation Studied by Brownian Dynamics Simulation, J. Colloid
Interface Sci., 231, 337-150 (2000)

\bibitem{Ref7}
H\"utter M., Brownian dynamics simulation of stable and of
coagulating colloids in aqueous suspension, Ph.D. thesis no.
13107, ETH Zurich, Switzerland (1999)

\bibitem{Ref8}
Russel, W. B., Saville, D. A. and Schowalter, W. R., Colloidal
Dispersions, Cambridge University Press (March 1989)

\bibitem{Ref9}
Wyss, H. M., H\"utter, M., M\"uller, M., Meier, L. P. and
Gauckler, L. J., Quantification of Microstructures in Stable and
Gelated Suspensions from Cryo-SEM, J. Colloid Interface Sci., 248,
340-346 (2002)

\bibitem{Ref10}
Cundall, P. A. and Strack, O. D. L., A discrete numerical model
for granular assemblies, G\'eotechnique, 29, 47-65 (1979)

\bibitem{Ref11}
PFC$^{\mathrm{3D}}$ User's Manual, Itasca Consulting Group,
Inc., Minneapolis, Minnesota, USA (1995)

\bibitem{Ref12}
Brown, E. T., Analytical and Computational Methods in Engineering
Rock Mechanics, Ed. London: Allen \& Unwin (1987)

\bibitem{Ref13}
Jaeger, H. M. and Nagel, S. R., Physics of the Granular State,
Science, 255 [\bf5051\rm], 1523-1531 (1992)

\bibitem{Ref14}
Wyss, H. M., Tervoort, E. V. and Gauckler, L. J., Mechanics and
Microstructures of Concentrated Particle Gels, J. Am. Ceram. Soc.,
88 [\bf9\rm], 2337-2348 (2005)

\bibitem{Ref15}
Kruyt, N. P. and Rothenburg, L., Statistics of the elastic
behaviour of granular materials, Int. J. Solids Struct., 38,
4879-4899 (2001)

\bibitem{Ref16}
O'Hern, C. S., Silbert, L. E., Liu, A. J. and Nagel, S. R.,
Jamming at zero temperature and zero applied stress: The epitome
of disorder, Phys. Rev. E, 68, 011306 (2003)

\bibitem{Ref17}
Zhang, H. P. and Makse, H. A., Jamming transition in emulsions and
granular materials, Phys. Rev. E, 72, 011301 (2005)

\bibitem{Ref18}
Crocker, J. C. and Grier, D. G., Methods of Digital Video
Microscopy for Colloidal Studies, J. Colloid Interface Sci., 179,
298-310 (1996)

\bibitem{Ref19}
Clarke, A. S. and J\'{o}nsson, H., Structural changes accompanying
densification of random hard-sphere packings, Phys. Rev. E, 47
[\bf6\rm], 3975-3984 (1993)

\bibitem{Ref20}
Schenker, I., Filser, F.T., Aste, T. and Gauckler, L.J.,
Microstructures and mechanical properties of dense particle gels:
Microstructural characterization, J. Eur. Ceram. Soc, 28
[\bf7\rm], 1443-1449 (2008)

\end{thebibliography}
\end{document}